\documentclass[11pt]{article}
\usepackage{moriond,epsfig,graphicx}

\def\be{\begin{equation}}
\def\ee{\end{equation}}
\def\bea{\begin{eqnarray}}
\def\eea{\end{eqnarray}}

\begin{document}
\vspace*{4cm}
\title{STATUS OF CKM ANGLE MEASUREMENTS, \\A REPORT FROM BABAR AND BELLE}

\author{ Owen Long }

\address{Department of Physics and Astronomy, University of California, \\ Riverside CA 92521, USA }

\maketitle\abstracts{
I will review the latest developments in determining the
$CP$-violating phases of the CKM matrix elements from measurements
by the BaBar and BELLE experiments at the high-luminosity $B$ factories
(PEP-II and KEKB).  The emphasis will be on the angle $\gamma/\phi_3$
of the Unitarity Triangle, which is the relative phase
$\arg(-V_{ud} V_{ub}^*/V_{cd} V_{cb}^*)$, or the $CP$-violating phase of the
$b\to u$ transition in the commonly used Wolfenstein convention.
}

\section{Introduction}

Only 8 years after the experimental discovery of $CP$ violation~\cite{cpv},
Kobayashi and Maskawa noted in a seminal paper~\cite{km} that extending
the quark sector to 3 generations would naturally introduce a
$CP$ violating phase in weak interactions.
The BaBar and BELLE experiments and the high-luminosity $B$ factories
(PEP-II and KEKB) at the SLAC National Accelerator Laboratory and
KEK were designed and built with the primary goal of performing the
first precision tests of the Kobayashi-Maskawa theory using $CP$ asymmetry
measurements in $B$ decays.
The unitarity constraint involving
the $1^{\rm st}$ and $3^{\rm rd}$ columns of the CKM quark mixing matrix
$V_{ud}V_{ub}^* + V_{cd}V_{cb}^* + V_{td}V_{tb}^* = 0$ is often visualized
as a triangle (``The Unitarity Triangle'') in the complex plane.
The $CP$ asymmetry measurements from BaBar and BELLE can be directly related
to the interior angles of the Unitarity Triangle with little theoretical
uncertainty~\cite{kirkbynir}.

\begin{figure}[ht!]
\begin{center}
\vspace{5mm}
\includegraphics[height=60mm]{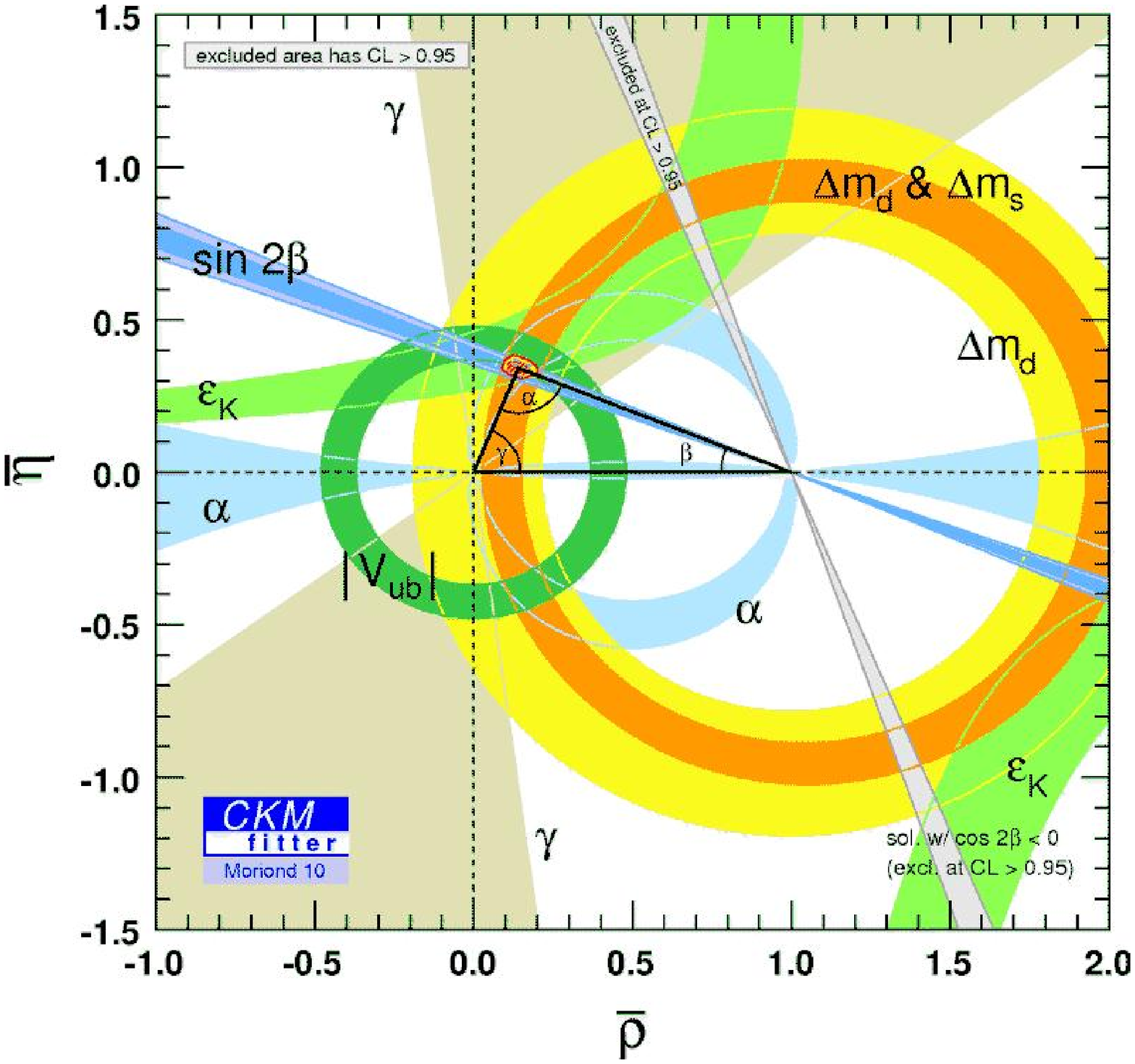}
\hspace{1cm}
\includegraphics[height=60mm]{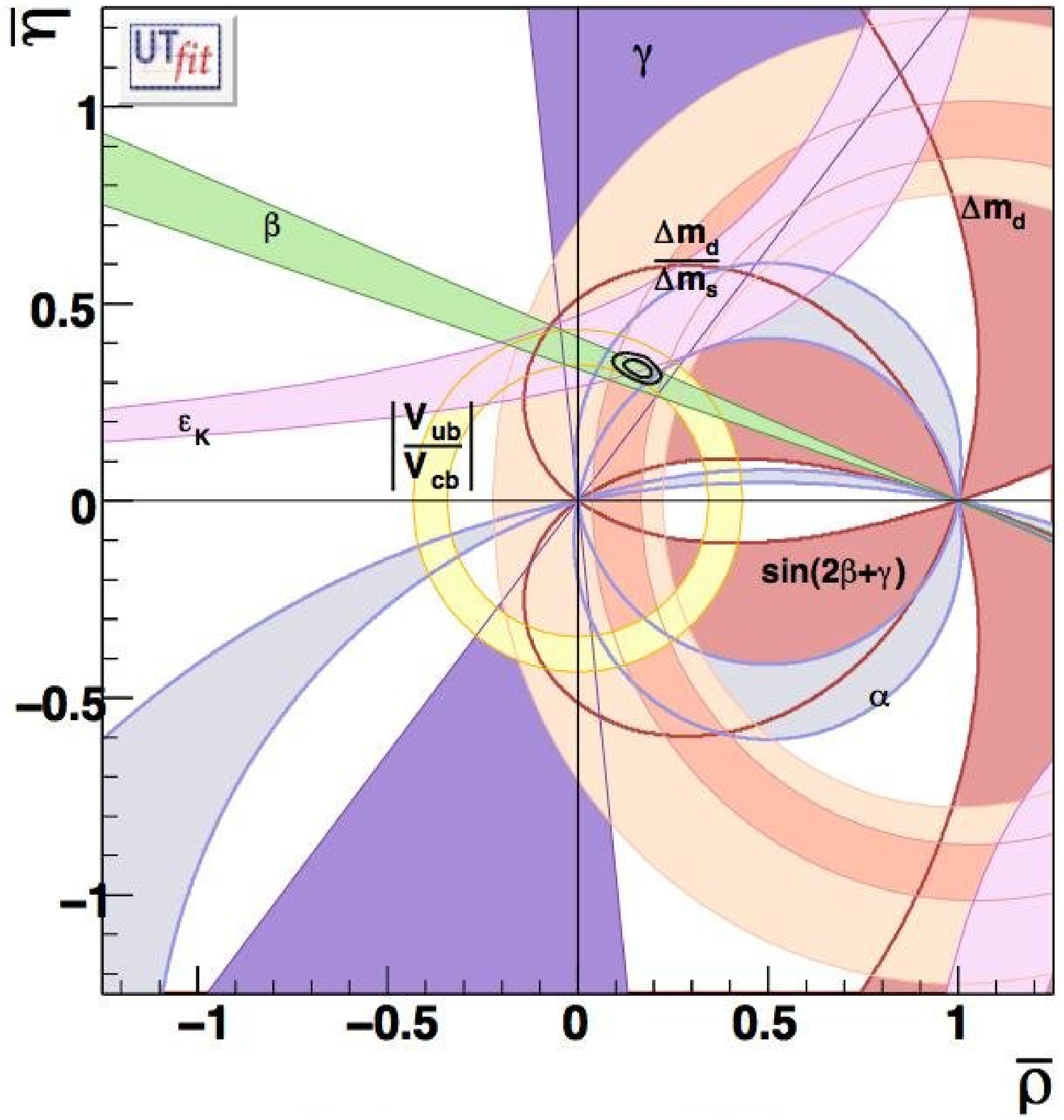}
\caption{ Experimental constraints on the Wolfenstein parameters $\bar \rho$ and $\bar \eta$.
           The 
           CKMfitter
           group (left) uses a frequentist statistical approach,
           while the 
           UTfit
           group (right) uses a Bayesian statistical approach.
}
 \label{fig:rhoeta}
\end{center}
\end{figure}

The current experimental constraints on the Wolfenstein parameters
$\bar \rho$ and $\bar \eta$, which give the coordinates of the tip
of the rescaled Unitarity Triangle in the complex plane, are shown
in Figure~\ref{fig:rhoeta}.
The analysis was done by two independent groups using different
statistical approaches (frequentist for CKMfitter~\cite{ckmfitter}
and Bayesian for UTfit~\cite{utfit}).
However, the conclusions are the same -- the $CP$ violation parameters
of the CKM matrix are overconstrained and
the Kobayashi-Maskawa theory has been experimentally confirmed.
Kobayashi and Maskawa were awarded half of the 2008 Nobel Prize in physics.

The constraint on the angle $\beta$ (or $\phi_1$), from the amplitude of
the proper-time-dependent $CP$ asymmetry of $B^0 \to J/\psi K^0_S$ and
other $b\to c\bar c s$ decays,
is the strongest, with a one standard deviation uncertainty of less than one
degree.
The most difficult angle to measure is $\gamma$ (or $\phi_3$).
Recent progress has been made over the past year in improving our measurements of $\gamma$
and I will focus on this for the rest of this writeup.

\section{ Methods for measuring $\gamma$ (or $\phi_3$) }

The angle $\gamma \equiv \arg(-V_{ud} V_{ub}^*/V_{cd} V_{cb}^*)$ can be
measured from direct $CP$ violation in $B$ decays where both $b\to c$ and $b\to u$
decay amplitudes contribute to the same final state and interfere with each other.
The methods~\cite{glw,ads,ggsz} that currently give the strongest constraints on
$\gamma$ use decays of the type $B^- \to D^0 K^-$ (from a $b\to c \bar u s$ decay)
with $B^- \to \bar D^0 K^-$ (from a $b \to u \bar c s$ decay) where the
$D$ decays to a final state that is accessible from both the $D^0$ and
the $\bar D^{0}$.
These are both tree level $b$ decays, so the interpretation of the measurements
in terms of $\gamma$ is theoretically extremely clean.
However, the ratio of the hadronic $B$ decay amplitudes $r_b \equiv |A(b\to u)/A(b\to c)|$
and the $CP$-conserving (strong) phase difference $\delta_b$ between
$A(b\to u)$ and $A(b\to c)$ can not be calculated with precision and must be experimentally
determined.
In addition to $\gamma$, all of the various $B\to DK$ methods share the same
hadronic parameters ($r_b$ and $\delta_b$).
Decays of the type $B\to D^*K$ and $B\to DK^*$ may also be used with each distinct $B$
decay having its own $r_b$ and $\delta_b$.

The precision of the current $\gamma$ measurements is limited due to two factors.
First, the signal samples of $B\to DK$ are relatively small (at most 100's of events) due to
CKM suppression of the decay amplitudes.
The second factor is that $r_b$ is relatively small (about 0.10 due to CKM and
color suppression) which limits the size
of the interference that we are trying to measure.

\section{ The $B\to DK$, $D$ decay Dalitz approach }

The best individual measurements of $\gamma$ come from using a 3-body $D$
decay (either $K^0_S \pi^+ \pi^-$ or $K^0_S K^+ K^-$) in the $B\to DK$ method.
The amplitude
(${\cal A_D}$)
and $CP$-conserving phase of the $D\to K^0_S h^+ h^-$ decay
varies accross the
$D$ decay Dalitz plot, which is the decay intensity in the plane of
$s^+=m^2(Kh^+)$ vs $s^-=m^2(Kh^-)$.
Assuming no $CP$ violation in $D$ decays, the $\bar D^0$ Dalitz plot is
the same as the the $D^0$ Dalitz plot after
reflection through the
$s^+ = s^-$
diagonal,
{\it i.e.} ${\cal A}_{\bar D}(s^+,s^-) = {\cal A}_{D}(s^-,s^+)$.
The parameters of a $D$ decay Dalitz amplitude model are determined from the data
by fitting a very clean, high statistics sample of flavor-tagged $D^0$ mesons
from $D^{*+}\to D^0\pi^+$ decays produced in $e^+e^-\to c\bar c$ events.
The overall amplitudes for the processes $B^\pm \to DK^\pm $; $D\to K^0_S h^+ h^-$   are given by
\begin{eqnarray}
\label{eqn:ampl1}
  A(B^+;s^+,s^-) & \propto & {\cal A}_{D}(s^-,s^+) \ + \ r_b \, e^{+i\gamma+i\delta_b} \ {\cal A}_{D}(s^+,s^-) \\
\label{eqn:ampl2}
  A(B^-;s^+,s^-) & \propto & {\cal A}_{D}(s^+,s^-) \ + \ r_b \, e^{-i\gamma+i\delta_b} \ {\cal A}_{D}(s^-,s^+)
\end{eqnarray}
where the first term is from the $b\to c$ transition and the second is from the $b\to u$ transition.
The relative weight of the two terms, both in magnitude and $CP$-conserving phase, is known
from ${\cal A}_{D}(s^+,s^-)$, apart from an overall factor of $r_b\, e^{\pm i\gamma+i\delta_b}$
that is experimentally determined in the data analysis.

\subsection{ The $B\to DK$, $D$ decay Dalitz data samples }

Both BaBar and BELLE have shown updates to their $\gamma$ measurements using the
$D$ decay Dalitz technique in the past year.
The BaBar collaboration has analyzed their full dataset, which contains
468 million $B\bar B$ events~\cite{babar-hepex}, while the Belle collaboration has shown
results using 657 million $B\bar B$ events~\cite{belle-hepex}.
The BaBar results with their full dataset were shown for the first time in this
talk.
Both the BaBar and BELLE analyses have been submitted for publication and are
still preliminary.
Both experiments have done the analysis for the following three $B$ decays:
$B^\pm \to DK^\pm$, $B^\pm \to D^*(D^0\pi^0,D^0\gamma)K^\pm$, and
$B^\pm \to D K^{*\pm}(K^0_S\pi^\pm)$
using the $D\to K^0_S \pi^+\pi^-$ decay mode.
The BaBar analysis also includes results using $D\to K^0_S K^+ K^-$.

The signal is separated from combinatoric background using two standard
reconstruction variables in the center of mass frame:
$m_{\rm ES} = \sqrt{ E^2_{\rm beam} - p^2_B}$ and
$\Delta E = E_B - E_{\rm beam}$.
Continuum ($e^+e^-\to q\bar q$) background is rejected using event shape variables
that are combined in an optimal linear combination (Fisher discriminant).
These shape variables take advantage of the fact that the decay products in
$B\bar B$ events are fairly isotropic, while continuum events have a preferred
direction along the $q\bar q$ axis.
Large $B^+ \to D^{(*)} \pi^+$ data control samples, where the $b\to u$ amplitude is more
suppressed with respect to the $b\to c$ transition ($r_b \approx 0.01$),
are used to calibrate and validate the analysis methods.

The Dalitz model parameters are determined from large, clean, flavor-tagged charm
samples from continuum production.
The $D^0\to K^0_S \pi^+\pi^-$ Dalitz models in the BaBar and BELLE analyses are not the same.
The main differences are in the treatment of the S-wave components.
%
Babar uses a K-matrix formalism with the P-vector approximation and 5 poles for
the $\pi\pi$ S-wave and
a LASS model consisting of a $K^*_0(1430)^\mp$ resonance together with a coherent
non-resonant contribution parameterized by a scattering length and an effective
range for the $K\pi$ S-wave.
BELLE includes $\sigma_1$ and $\sigma_2$ $\pi\pi$ scalar resonances and
a $K^*_0(1430)$ for the $K\pi$ S-wave.
Details of the Dalitz models can be found in the preprints~\cite{babar-hepex,belle-hepex} describing
the measurements.

Table~\ref{tab:yields} gives the $B\to D^{(*)}K^{(*)}$ signal yield for the samples
used in the final fits for the $CP$ parameters (described below).
The BaBar signal efficiencies have improved substantially (20\% to 40\% relative)
with respect to the previous BaBar analysis, which used 383 million $B\bar B$ events~\cite{babar-oldgamma},
coming mainly from reprocessing the data with improved track reconstruction and particle
identification.

\begin{table}
  \begin{center}
     \begin{tabular}{|l||c|c|c|}
       \hline
         $B$ decay mode  &  BELLE ($K^0_S\pi^+\pi^-$)  & BaBar ($K^0_S\pi^+\pi^-$) & BaBar ($K^0_SK^+K^-$) \\
                         &  657 M $B\bar B$            &  468 M $B\bar B$  &  468 M $B\bar B$ \\
       \hline \hline
       $B^\pm \to D K^\pm$             &  $757 \pm 30$  &  $920 \pm 35$  &  $142 \pm 14$  \\
       $B^\pm \to D^*(D\pi^0) K^\pm$   &  $168 \pm 15$  &  $246 \pm 22$  &   $53 \pm 11$  \\
       $B^\pm \to D^*(D\gamma) K^\pm$  &   $83 \pm 10$  &  $191 \pm 19$  &   $31 \pm  7$  \\
       $B^\pm \to D K^{*\pm}$          &                &  $163 \pm 17$  &   $28 \pm  6$  \\
       \hline
     \end{tabular}
     \caption{ Signal yields for the samples used in the final fits for the $CP$ parameters. These
                results are preliminary.}
     \label{tab:yields}
  \end{center}
\end{table}

\subsection{ The $B\to DK$, $D$ decay Dalitz $CP$ analysis }

The $CP$ parameters are determined using unbinned maximum likelihood fits.
Probability density functions in the likelihood depend on $\Delta E$, $m_{\rm ES}$,
continuum rejection variables, and the Dalitz plot position.
The interference terms in the intensity are proportional to
\begin{equation}
  x_\pm = r_b \, \cos(\delta_b \pm \gamma) \ \ \ \ {\rm and}  \ \ \ \ y_\pm = r_b \, \sin(\delta_b \pm \gamma).
\end{equation}
The Cartesian parameters are free parameters in the fits.
They are used rather than $r_b$, $\delta_b$, and $\gamma$ directly because they
are uncorrelated with Gaussian uncertainties.

The full results of the fits can be found in the BaBar and BELLE preprints~\cite{babar-hepex,belle-hepex}
and averages are available through HFAG~\cite{hfag}.
Table~\ref{tab:xy} gives the $x_\pm$ and $y_\pm$ results for the $B^+\to DK^+$
mode as an example to give you an idea of the measurement precision and the consistency of the two
measurements.
The BaBar and BELLE results are consistent with each other.
The BaBar statistical errors are lower due to the higher signal statistics (see Table~\ref{tab:yields}).
The degree to which the $x_\pm$ and $y_\pm$ are inconsistent with zero is the significance of the
$b\to u$ transition, while the degree to which $x_- \neq x_+$ and $y_- \neq y_+$ is the significance
of the $CP$ violation.

\begin{table}
  \begin{center}
    \begin{tabular}{|c||r|r|}
      \hline
         \multicolumn{3}{|c|}{ $CP$ parameters for $B^+\to DK^+$ , $D\to K^0_S\pi^+\pi^-$ } \\
         \hline
         Parameter  &  \multicolumn{1}{|c|}{BaBar}  &  \multicolumn{1}{|c|}{BELLE} \\
         \hline\hline
          $x_-$ (\%)  &   $6.0 \pm 3.9 \pm 0.7 \pm 0.6$   &   $10.5 \pm 4.7 \pm 1.1$  \\
          $y_-$ (\%)  &   $6.2 \pm 4.5 \pm 0.4 \pm 0.6$   &   $17.7 \pm 6.0 \pm 1.8$  \\
          $x_+$ (\%)  & $-10.3 \pm 3.7 \pm 0.6 \pm 0.7$   &  $-10.7 \pm 4.3 \pm 1.1$  \\
          $y_+$ (\%)  &  $-2.1 \pm 4.8 \pm 0.4 \pm 0.9$   &   $-6.7 \pm 5.9 \pm 1.8$ \\
      \hline
    \end{tabular}
  \end{center}
  \caption{Preliminary results of the $CP$ fit for $B^+\to DK^+$.  The
      BELLE fit uses $D\to K^0_S\pi^+\pi^-$, while the BaBar fit uses
      both $D\to K^0_S\pi^+\pi^-$ and $D\to K^0_S K^+K^-$.  The uncertainties
       from left to right are statistical, experimental systematic, and Dalitz model systematic.
       The BELLE analysis does not report Dalitz model uncertainties on $x_\pm$ and $y_\pm$.}
  \label{tab:xy}
\end{table}

The interpretation of selected $x_\pm$ and $y_\pm$ measurements is given in
Table~\ref{tab:rbgamma}.
Each experiment independently finds a value of $\gamma$ close to around $70^\circ$,
which is consistent with indirect determinations of $\gamma$ within the $CKM$ framework
(see Figure~\ref{fig:rhoeta} and refs~\cite{ckmfitter,utfit}).
Each experiment rules out $CP$ conservation with a significance of 3.5 standard deviations.
The Belle experiment favors a larger $b\to u$ contribution to the decay (larger $r_b$),
which leads to a smaller statistical uncertainty on $\gamma$, though the BELLE
and BaBar $r_b$ measurements are not incompatible.
Both the BaBar and BELLE measurements are statistics limited.

One noteworthy difference between the BaBar and BELLE measurements is the uncertainty
from the Dalitz model, which is $3^\circ$ for BaBar and $8.9^\circ$ for BELLE.
The BaBar $\gamma$ analysis~\cite{babar-hepex} and $D^0$ mixing analysis~\cite{babar-d0mix}
used the same Dalitz model and the same model variations in the evaluation of the
systematic uncertainties.
The Dalitz model systematic errors are not negligible in the $D^0$ mixing analysis, so the
Dalitz model and model variations were refined and reconsidered, with respect to the initial BaBar $D$ decay Dalitz $\gamma$
analysis~\cite{babar-oldgamma}.
This Dalitz model work, motivated by the requirements of the $D^0$ mixing analysis, was propagated
back into the $\gamma$ analysis, which lead to the substantial improvement in the model systematic
uncertainty on $\gamma$.
In the future, LHCb and super $B$ factories will have much larger datasets, making model
independent approaches~\cite{ggsz,model-independent,cleo} feasible.

\begin{table}
  \begin{center}
    \begin{tabular}{|c||c|c|}
      \hline
         Parameter  &  \multicolumn{1}{|c|}{BaBar}  &  \multicolumn{1}{|c|}{BELLE} \\
         \hline\hline
         $\gamma \ (^\circ)$           &  $68\ ^{+15}_{-14} \ \{4,3\}$            &   $78.4\ ^{+10.8}_{-11.6} \ \pm 3.6 \ \pm 8.9$   \rule{0pt}{5mm} \\
         $r_b$, $DK \ (\%)$            &  $9.6 \pm 2.9 \ \{0.5, 0.4\}$            &   $16.0\ ^{+4.0}_{-3.8}   \ \pm 1.1 \ ^{+5.0}_{-1.0}$     \rule{0pt}{5mm} \\
         $r_b^*$, $D^*K \ (\%)$          &  $13.3 \ ^{+4.2}_{-3.9} \ \{1.3, 0.3\}$  &   $19.6\ ^{+7.2}_{-6.9}\ \pm 1.2 \ ^{+6.2}_{-1.2}$     \rule{0pt}{5mm} \\
         $\delta_b$, $DK\ (^\circ)$    &  $119\ ^{+19}_{-20} \ \{3,3\}$           &   $136.7\ ^{+13.0}_{-15.8}\ \pm 4.0 \ \pm 22.9$  \rule{0pt}{5mm} \\
         $\delta_b^*$, $D^*K\ (^\circ)$   &  $-82 \pm 21\ \{5,3\}$                  &   $341.9\ ^{+18.0}_{-19.6}\ \pm 3.0 \ \pm 22.9$  \rule{0pt}{5mm} \\
      \hline
    \end{tabular}
  \end{center}
  \caption{ Interpretation of selected $x_\pm$ and $y_\pm$ measurements.  For the BaBar results,
              the first uncertainty gives the 68.3\% confidence interval including all sources of
              uncertainty (stat., expt. syst., Dalitz model).  The values inside the \{\}
              indicate the symmetric contributions to the total uncertainty coming from the
              experimental systematic and Dalitz amplitude model systematic uncertainties, respectively.
              For the BELLE results, the uncertainties from left to right are statistical, systematic, and Dalitz model
              systematic.  All results are preliminary.
              }
  \label{tab:rbgamma}
\end{table}

\section{ The ``ADS'' approach for $\gamma$}

The so-called ``ADS'' method (for Atwood, Dunietz, and Soni~\cite{ads}) for determining
$\gamma$ maximizes the size of the interference term with a clever choice of final state.
The favored $b\to c$ transition from $B^- \to D^0 K^-$ is combined with the suppressed
$c\to d$ transition from $D^0\to K^+\pi^-$.
This interferes with the suppressed $b\to u$ transition from $B^- \to \bar D^0 K^-$
followed by the favored $\bar c \to \bar s$ transition from $\bar D^0 \to K^+\pi^-$.
Since both paths to the $[K^+\pi^-]_D K^-$ final state involve a CKM favored transition
combined with a CKM suppressed transition, the paths have roughly equal amplitudes.
This means the direct $CP$ asymmetry can be quite large (of order 1) but you pay a
heavy price in signal statistics due to the CKM suppression.
This method is quite sensitive to the amplitude ratio $r_b$, which is common with the
other $B\to DK$ methods, such as the $D$ decay Dalitz method above.

\begin{figure}[h!]
\begin{center}
\vspace{5mm}
\includegraphics[height=50mm]{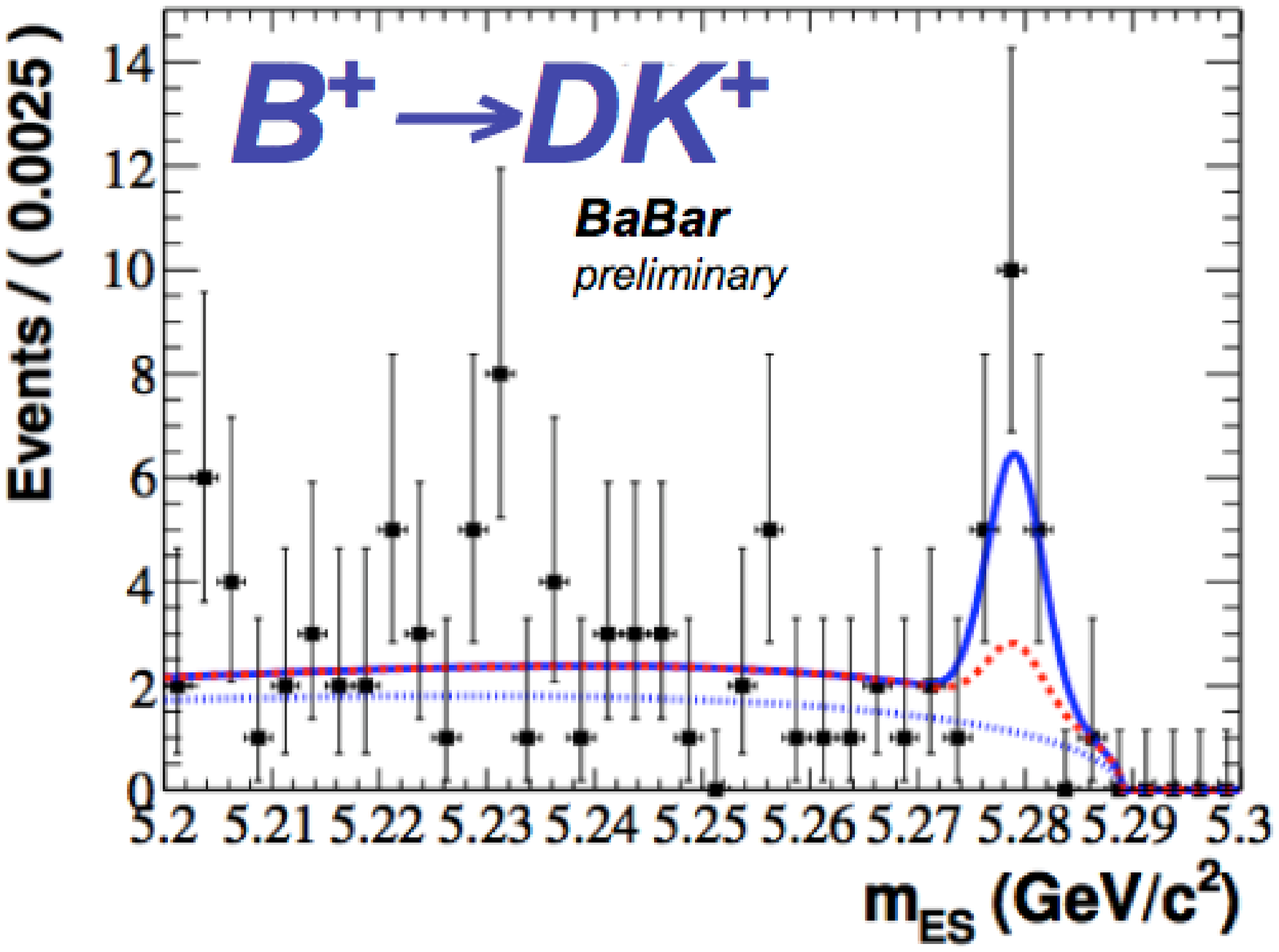}
 \hspace{1cm}
\includegraphics[height=50mm]{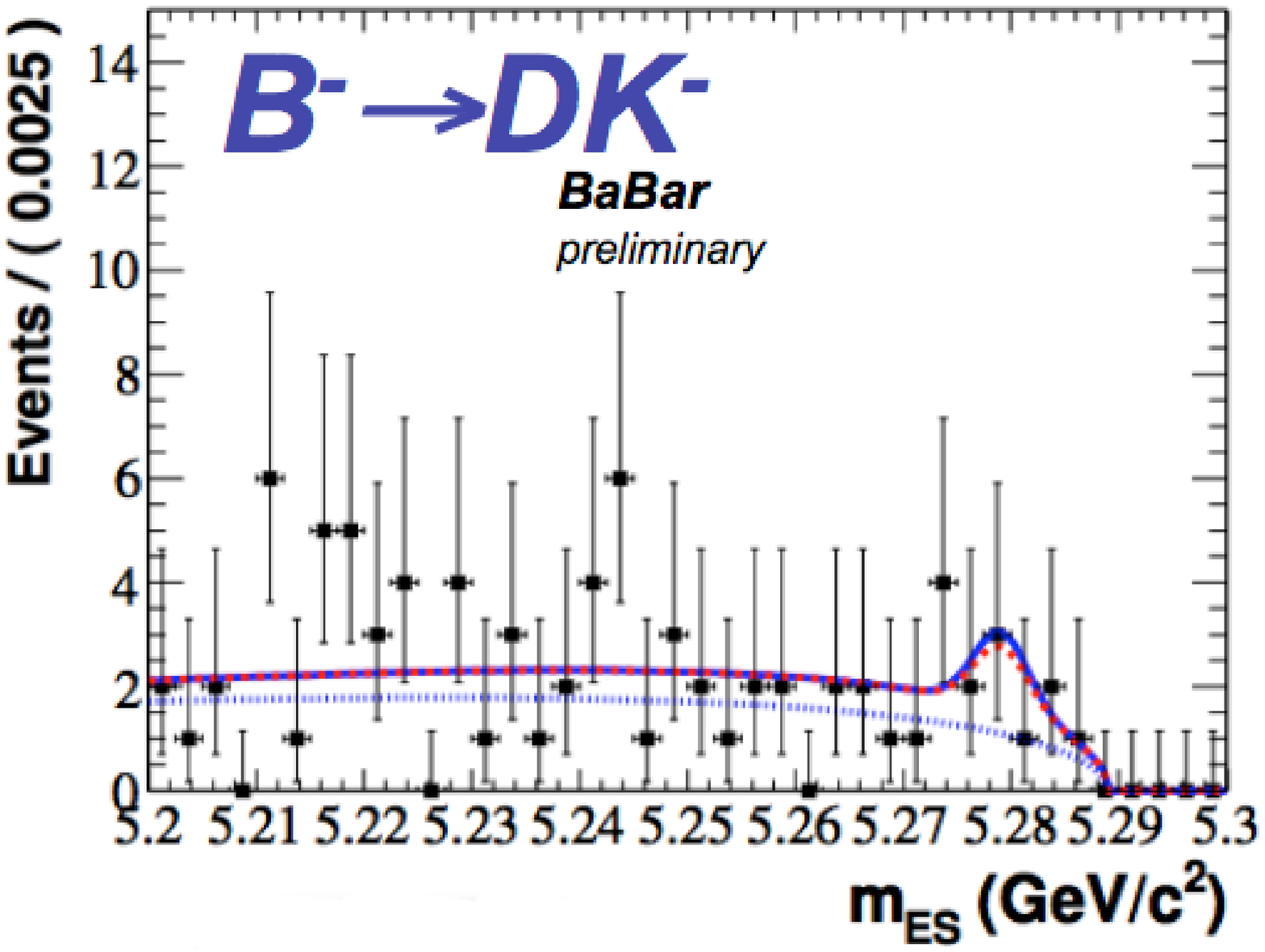}
\caption{ Distributions of $m_{\rm ES}$ for $B^+ \to [K^-\pi^+]_D K^+$ (left)
   and $B^- \to [K^+\pi^-]_D K^-$ (right) for the preliminary BaBar analysis
   of 468 million $B\bar B$ events.  The dotted blue curve is combinatoric background
   only, the dashed red curve is combinatoric plus peaking background, and the
   solid blue curve represents all components including the signal.
}
 \label{fig:ads}
\end{center}
\end{figure}

Both BaBar and BELLE have searched for $B^- \to [K^+\pi^-]_D K^-$.
The BaBar collaboration recently released a preliminary 
version of their analysis using the full
dataset of 468 million $B\bar B$ events.
Unlike previous searches from both experiments, the new BaBar analysis sees
the first signs of ADS signals in $B^\pm \to DK^\pm$ and $B^\pm \to D^*K^\pm$.
Figure~\ref{fig:ads} shows the $m_{\rm ES}$ distributions separately for
$B^+ \to [K^-\pi^+]_D K^+$ and $B^- \to [K^+\pi^-]_D K^-$.
Comparing the $B^+$ and $B^-$ distributions, a large $CP$ asymmetry is evident.

The interpretation of the ADS rate and $CP$ asymmetry gives
$r_b = (9.0^{+5.6}_{-5.1})\%$ and $r^*_b = (11.6^{+3.4}_{-5.0})\%$ for $B\to DK$
and $B\to D^*K$ respectively and constraints on $\gamma$, $\delta_b$, and $\delta_b^*$
that are consistent with the $D$ decay Dalitz measurements.

\section{ Summary and future prospects }

The CKM parameters are now over constrained.
All $CP$ violation measurements made thus far are consistent with the
Kobayashi-Maskawa mechanism of the Standard Model.
The $B$ factory experiments, BaBar and BELLE, have made recent progress on
the most difficult Unitarity Triangle angle to measure: $\gamma$ or $\phi_3$.
Unlike other angle measurements, $\gamma$ from $B\to DK$ involves only tree-level
processes, which make the interpretation very clean theoretically, providing
a solid Standard Model reference.
However, our experimental constraints on $\gamma$ from $B\to DK$ are still
relatively weak and statistics limited.
The analysis of all experimental constraints by the UTfit~\cite{utfit}
and CKMfitter~\cite{ckmfitter} collaborations
gives 
$\gamma = \left( 72 \pm 11\right)^\circ$
and
$\gamma = \left( 69\, ^{+19}_{-21} \right)^\circ$, respectively.

Looking ahead, the LHCb experiment will make substantial progress on
$\gamma$ using $B\to DK$ decays, taking advantage of
the huge $b\bar b$ production cross section in $pp$ collisions to
address the current limitation, which is signal statistics.
The high statistics will make model-independent $D$ decay Dalitz approaches
viable, removing the dependence on the Dalitz amplitude model assumptions.
%
%
A super $B$ factory could also turn $\gamma$ into a precision measurement.


\section*{Acknowledgments}
I would like to thank Fernando Martinez-Vidal and Anton Poluektov for providing details
of the BaBar and BELLE $B\to DK$, $D$ decay Dalitz analyses and advice on how to best present
the measurements, Tim Gershon for the HFAG averages,
and Vincent Tisserand and Achille Stocchi for providing the CKMfitter and UTfit analysis
results.  I would also like to thank Alex Bondar for some interesting conversations in LaThuile
about the material in these proceedings.

\end{document}